\documentclass[bimj,fleqn]{w-art}
\usepackage{times}
\usepackage{changebar}
\usepackage{w-thm}
\usepackage[authoryear]{natbib}
\usepackage{graphicx, bm, url}
\usepackage{amsmath,amsfonts}
\usepackage{natbib}

\usepackage{xcolor}         

\definecolor{DarkGreen}{rgb}{0.5,0.8,0.6}   
\definecolor{RGBblack}{rgb}{0.0,0.0,0.0}    

\newcommand{\TT}{{\mbox{TT}}}
\newcommand{\Oth}{O}
\newcommand{\As}{A^\star}

\newcommand{\yb}{\bm{y}}

\newcommand{\xb}{\bm{x}}
\newcommand{\Xb}{\bm{X}}
\newcommand{\xbs}{\bm{x^\star}}
\newcommand{\xis}{\xi^\star}
\newcommand{\xibs}{\bm{\xi^\star}}

\newcommand{\mb}{\bm{m}}

\newcommand{\balpha}{\bm{\alpha}}

\newcommand{\xits}{\widetilde{\xi}^{\star}}

\newcommand{\bth}{\bm{\theta}}
\newcommand{\bths}{\bm{\theta^\star}}

\newcommand{\Dir}{\text{Dir}}

\newcommand{\mult}{\text{mult}}
\newcommand{\LN}{\text{LN}}
\newcommand{\N}{\text{N}}

\renewcommand{\th}{\theta}

\newcommand{\sig}{\sigma}

\newcommand{\HR}{\mbox{HR}}
\newcommand{\NA}{\mbox{NA}}
\newcommand{\AH}{\mbox{AH}}

\newcommand{\db}{A} 
\newcommand{\dbtrue}{A^{\mbox{true}}}

\newcommand{\TE}{\mbox{TE}}
\newcommand{\TEhat}{\widehat{\TE}}
\newcommand*\rot{\footnotesize\rotatebox{90}}

\theoremstyle{plain}

\theoremstyle{definition}

\usepackage[]{graphicx}
\chardef\bslash=`\\ 

\begin{document}
\DOIsuffix{bimj.200100000}
\Volume{52}
\Issue{61}
\Year{2017}
\pagespan{1}{}

\title[A Nonparametric Bayesian
  Basket Trial Design]{A Nonparametric Bayesian
  Basket Trial Design}
\author[Xu {\it{et al.}}]{Yanxun Xu\footnote{Corresponding author: {\sf{e-mail: yanxun.xu@jhu.edu}}, Phone: +1-410-516-7341}\inst{,1}} 
\address[\inst{1}]{Department of Applied Mathematics and
    Statistics, Johns Hopkins University, Baltimore, MD, 21218, USA}
\author[]{Peter M\"uller\inst{2}}
\address[\inst{2}]{Department of Mathematics, University of Texas at
    Austin, Austin, TX, 78705, USA}
\author[]{Apostolia M Tsimberidou\inst{3}} \address[\inst{3}]{Department of  Investigational Cancer Therapeutics, The
    University of Texas M.D. Anderson Cancer Center, Houston, TX, 77005, USA}
  \author[]{Donald Berry\inst{4}} \address[\inst{4}]{Department of  Biostatistics, The
    University of Texas M.D. Anderson Cancer Center, Houston, TX, 77005, USA}
\Receiveddate{zzz} \Reviseddate{zzz} \Accepteddate{zzz} 

\begin{abstract}
Targeted therapies on the basis of genomic aberrations analysis of the
tumor have shown promising results in 
cancer prognosis and
treatment.  Regardless of tumor type, trials that match patients to
targeted therapies for their particular genomic aberrations  
have become a mainstream direction of therapeutic management of
patients with cancer. 
Therefore, finding the subpopulation of patients who can most benefit
from an aberration-specific targeted therapy across multiple cancer
types is important.  We propose an adaptive Bayesian clinical trial
design for patient allocation and subpopulation identification. We start
with a decision theoretic approach, including a utility function
and a probability model across all possible subpopulation models. The main
features of the proposed design and population finding methods are
the use of a flexible non-parametric Bayesian survival regression based on a
random covariate-dependent partition of patients,
and decisions based on a flexible utility function that reflects the
requirement of the clinicians appropriately and realistically, and the
adaptive allocation of patients to their superior treatments. Through
extensive simulation studies, the new method is demonstrated to achieve desirable operating characteristics and compares favorably against the alternatives. 

\end{abstract}

\maketitle                   






\section{Introduction}

We propose an adaptive Bayesian clinical trial design for patient
allocation and subpopulation finding in a heterogeneous patient
population in basket trials.  We focus on the objectives of
allocating patients to their superior treatments and
identifying a subpopulation of patients who are most likely to benefit
from the targeted therapy under consideration.

Recent developments of genomic profiling technologies 
\citep{Snij,Vijver:2002,  barskigenomic2009,
baladandayuthapanibayesian2010, curtisthe2012,
xu2013nonparametric}
have revolutionized the traditional diagnosis and treatment of cancer,
leading to the development of targeted therapies designed to
target specific biomarkers and molecular pathways involved in the
pathophysiology of tumor initiation, metastasis, and drug resistance.
For example, matching genomic aberrations with targeted therapies has
led to 
the use of trastuzumab on HER2+ breast cancer
\citep{hudis2007trastuzumab}, and the recommendation against EGFR
antibodies therapy for KRAS mutated colorectal cancer
\citep{misale2012emergence}.

Some studies investigate the matching of
tumor molecular alterations regardless of patient's tumor type. 
One of the first such trials was the IMPACT 
(Initiative for Molecular Profiling and Advanced Cancer Therapy)
study \citep{impact, tsimberidou:12} which investigated the use of targeted
agents matched with tumor molecular aberrations, and the following study
IMPACT II \citep{impact2} which we describe in this paper. 
Such trials are known as ``basket" trials. Similar later studies that followed this example include the
 Lung-MAP
(NCT021544490) and MATCH trials \citep{conley2014molecular}, which enrolled
patients into sub-studies based on their genomic alterations.  
Next-generation
sequencing (NGS) is used to identify patients with a specific genomic
alteration or mutation, regardless of the specific cancer. 
Patients are
then enrolled in a trial to assess a
particular molecularly targeted therapy.  For instance, BRAF is a
relatively common mutation in patients with melanoma, for 
which  Vemurafenib (Zelboraf)
was approved in 2011, but also occurs less frequently in other
types of cancer. Researchers found that the drug showed little
efficacy in patients with BRAF- mutant colorectal cancer
\citep{yang2012antitumor}. Therefore, finding the subpopulation of patients
who can benefit from a mutation-specific experimental therapy across
multiple tumor types is important to ``basket" trials. For example, \cite{hyman2015vemurafenib} systematically studied Vemurafenib in 122 patients with  non-melanoma cancers harboring BRAF mutation and showed that Vemurafenib had efficacy for patients with BRAF mutation in non-small-cell lung cancer and Erhheim-Chester disease.

There is a growing literature to propose Bayesian approaches to
identify subpopulations with enhanced treatment effects.
 The general problem of reporting exceptions to an overall
conclusion of a clinical study is known as subgroup analysis. 
\cite{dixon1991bayesian} approach subgroup analysis as inference
on
treatment/covariate interaction effects.
\cite{simon2002bayesian} uses a similar approach with
independent priors on the interaction
parameters. 
\cite{sivaganesan2011bayesian} consider subgroup analysis
as a model selection problem with each covariate defining a family of
models.  \cite{ruberg2010mean} and \cite{foster2011subgroup} develop
tree-based algorithms to identify and evaluate subgroup effects by
searching for regions with substantially enhanced treatment effects
compared to the average effect, averaging across the covariate
space. \cite{laud2013subgroup} report subgroups within a Bayesian
decision-theoretic framework.  They determine rules using an extension
of a 0/1/K utility function. The utility function is based on the
posterior odds of subgroup models relative to the overall null and
alternative models. \cite{xu2014subgroup} identify subgroups of patients with substantially different treatment effect based on a partition of the biomarker space using a variation of Bayesian classification and regression tree.  

Several recent clinical trials explore the use of Bayesian adaptive designs,
in combination with  subpopulation finding designs.
Prominent examples include the breast 
cancer trial ISPY-2 \citep{barker2009spy} that uses indicators for
several biomarkers and a MammaPrint risk score to define 14
subpopulations of possible practical interest. The design graduates
subpopulations, that is, recommends a future phase III study; or drops
subpopulations or treatment arms, that is, remove one of the 14
subpopulations or treatment arms from further consideration.  A
similar design is the BATTLE study of \cite{battle:08} who define 5
subpopulations of lung cancer patients based on biomarker profiles and
proceed to adaptively allocate patients to alternative treatments.
Another recent discussion is \cite{berryAl13} who include a 
comparison of Bayesian adaptive designs, including a design based on a
hierarchical model over different subpopulations with a comparable
design using Simon's optimal two-stage design \citep{simon12}.

Some recent frequentist approaches use Bayesian
methods to determine the adaptive enrichment to a subpopulation that
is most likely to benefit from a treatment 
\citep{SimonSimon:17}.
\cite{BrannathAl:09} use posterior predictive probabilities to propose
the adaptive enrichment in a seamless phase II/III design.
One of the challenges of such approaches is the control of
(frequentist) operating characteristics.
\cite{BretzAl:06}, in a seamless phase II/III design with a selection
among multiple candidate treatments (doses), 
achieve the desired analytic error control by using combination tests to
combine phase II and III data and closed testing to control for
multiplicities
(the use of combination tests for error control is not restricted to
the special case of selecting one treatment arm or dose, but is
fairly general across many different design modifications). 
Quantifying
errors and uncertainties for the more general problem of inference for
a benefiting subpopulation without pre-defined candidates is more
challenging.  \cite{Schnell:16} and \cite{Schnell:17} propose a
principled Bayesian approach by defining a notion of posterior
credible intervals for the estimated subpopulation.

In this paper, we  build on these earlier approaches and 
propose an adaptive Bayesian clinical trial design
for patient allocation in basket trials and a decision theoretic approach for subpopulation
finding in a heterogeneous patient population. 
Methodologically, we cast the problem as a
decision problem and separate the assumed sampling model and the
 decision problem. 
The important implication is that the description of the
desired subpopulation does not hinge on inference for parameters in
the sampling model, but is treated as a separate element in the
statistical inference problem.

To proceed, we first introduce the IMPACT II study in Section 2.
The proposed design is summarized in Section 3.
Next, in Section 4 we discuss the
subpopulation selection.
The discussion is possible without reference to details of the probability model. 
In Section 5 we introduce the particular survival regression model
that we use and adaptive treatment allocation.
Section 6 reports simulation studies. Finally, we conclude with a discussion in Section 7.

\section{IMPACT II}
\label{sec:mot} The proposed design is motivated by a
clinical trial, IMPACT II \citep{impact2,Tismb&Schilsky:14},
conducted at M.D. Anderson Cancer
Center, based on data from multiple tumor types and molecular aberrations. 
The primary objective of the study is to determine if
patients treated with
a targeted therapy (TT) that is selected based
on mutational analysis of the tumor have longer progression-free
survival than those treated with other therapy (\Oth). An important
secondary aim is to identify a subpopulation of patients who might
most benefit from targeted therapy. In other words, identify a
subpopulation that could best define eligibility criteria for a future
study of targeted therapy. Genomic analysis of tumor samples is
performed at the time of enrollment to identify tumor molecular
aberrations and to assign treatment for every individual patient.

\begin{figure} \centering
\rot{\hspace{3cm} FAILURE RATE, \%}
\includegraphics[width=.65\textwidth]{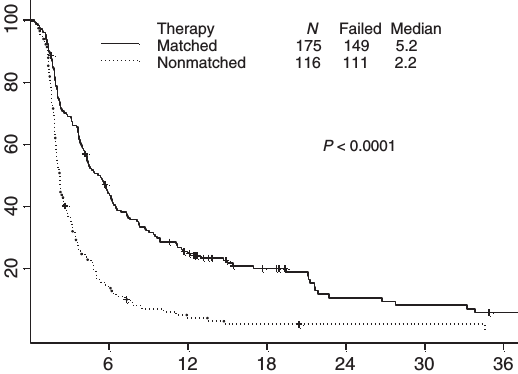}\\
MONTHS\\
\caption{Summary of observational, non-randomized data in IMPACT trial.  The plot
compares Kaplan Meier plots for PFS under TT versus \Oth.  The lack of
randomization prevents a causal interpretation.}
\label{fig:obs}
\end{figure}

Related observational data from a comparable patient population was 
reported in the IMPACT study
\citep{impact, tsimberidou:12, TsimbAlBerry:14}. 
This previous exploratory,
non-randomized study was also performed at M.D. Anderson Cancer Center
to test whether the use of targeted agents matches with tumor
molecular aberrations would improve clinical outcomes compared to the
standard approach. Figure \ref{fig:obs} summarizes the data from IMPACT as a
Kaplan-Meier plot comparing TT versus \Oth. The 
 plot indicates that
patients who were treated with TT based on  
their tumor molecular profiling 
 (labeled ``matched'' in the figure) 
could have superior clinical outcomes
compared to those who were treated with the conventional
approach (``non-matched''). However, this exploratory study was not
randomized. Unknown confounding factors may have
contributed to higher rates of response and longer time to treatment
failure and survival in patients treated with TT compared to those
treated with \Oth. One possible confounding factor could be a more
favorable prognosis for patients treated with TT, such as EGFR
mutation, which is well known to confer a more favorable prognosis.

To overcome the limitations of the previous exploratory study IMPACT and to
quantify the benefits of assigning therapy based on mutational
analysis over standard therapy, we use a randomized study for the new trial: 
IMPACT II. 
For each enrolled patient we record
a set of molecular aberrations,
$\mb_i=(m_{i1},\ldots,m_{iq})$ and tumor type $c_i$ and 
 we decide  a treatment allocation $z_i \in$
\{\Oth,TT\} for either 
 targeted therapy matched to a molecular aberration (TT), or
   other therapy, excluding targeted therapy (\Oth).
Importantly, the set of molecular aberrations that is recorded
for each patient can vary substantially. We  use $m_{ij}=\NA$ for
not recorded aberrations and $m_{ij} \in \{0,1\}$ for the absence or
presence of recorded aberrations, respectively.
Note that mutations are not mutually exclusive and any patient
could record multiple aberrations. When a patient has multiple aberrations that are eligible for targeted therapy and is allocated to TT, then the treatment is chosen based on an ordered list of mutations and drugs to treat the mutations (this list is established by the tumor board which is established as part of the protocol). So we denote $m_{ij}=1$ is aberration $j$ is targeted to treat for patient $i$, otherwise 0. 
Denote 
the combined covariate vector by $\xb_i = (\mb_i, c_i)$. Finally,
$y_i$ records progression free survival (PFS) time.

The proposed design is based on a survival regression for PFS $y_i$ as
a function of $\xb_i$ and $z_i$.  Continuously updated inference under
this model is used for adaptive treatment allocation during the trial.
Patients are assigned to TT or \Oth\, with probabilities that are related
to the predictive distribution of PFS under the two treatment arms.
At the end of the trial we use inference under the same model to
recommend a patient population for a future trial.

\section{Design}
In IMPACT II, patients with metastatic cancer (any tumor type)
and up to three prior therapies will undergo tumor biopsy followed by
molecular profiling.
For each patient we record tumor type and presence/absence of
a set of molecular aberrations, including
 PIK3CA, PTEN, BRAF, MET, and ``others'' (including but not limited to FGFR
alterations).
Patients with colorectal cancer and BRAF mutation were excluded 
because of available data demonstrating that these inhibitors have no
activity as single agents. 

 If at least one molecular alteration is
identified, the patient will be treated as follows: if there is a
U.S. Food and Drug Administration (FDA)-approved drug within the
labeled indication, the patient will receive it; if there is no
approved drug for the alteration and the tumor type, but there is a
commercially available targeted agent or appropriate clinical trial,
patients will be randomly selected to receive targeted therapy 
(TT)  versus
treatment not selected on the basis of genetic profiling  (O). 
The allocation probabilities for the random selection are specified as
follows in an initial run-in phase and a later adaptive allocation
phase of the trial.

\label{sec:design} \underline{Run-in phase.}  We initiate the study
with a run-in phase consisting of $n_0=100$  randomized 
patients. During this phase, patients who are selected for
randomization (as described above) 
are equally randomized to the two treatment arms TT vs. \Oth.  

\underline{Adaptive allocation.}  After the initial run-in
of 100 patients, we introduce adaptive randomization for the next
$n_1=300$ patients with a cohort size 50, allowing for a total of
$n_{\mbox{max}}=400$  randomized  patients.  For the adaptive randomization, we use
model-based posterior predictive probabilities. We discuss details of
the model specification later. Let $\pi_i$ denote the posterior
probability, based on current data, that PFS time under TT is greater
than survival under \Oth\, for patient $i$.  See \S5 for the
evaluation of $\pi_i$ in the implemented probability model.  Let $z_i
\in$ \{TT,\Oth\} denote the treatment allocation for patient $i$.  
 For patients who are selected for randomization (as described
above),  we use adaptive treatment allocation
\begin{equation}
  p(z_i = \TT) = \begin{cases}
    p_0   & \mbox{ if }  \pi_i < p_0 \\
    \pi_i & \mbox{ if } p_0 < \pi_i < p_1\\
    p_1   & \mbox{ if }  \pi_i >p_1.
   \end{cases}
\label{eq:alloc}
\end{equation}
We use $p_0=0.1$ and $p_1=0.9$.

\underline{Subpopulation finding.}
Most importantly, 
at the conclusion of the
trial (6 months after 400 patients accrue), we assess
subpopulation-specific effects of targeted therapies.  We report the
subpopulations that achieve the maximum benefit from targeted
therapies.  This assessment is based on the posterior expected utilities
of all possible subpopulations.  The utilities depend on the
characteristics of subpopulations as well as the posterior probability
model under an estimated regression for PFS.  We shall evaluate the
log hazards ratio with respect to PFS. This will be measured from the
time of initiation of treatment to disease progression or death, as of
the last follow-up visit.
Details of the subpopulation finding are described below, in \S
4. 

\section{Subpopulation Finding}
\label{sec:subpop}
Recall that $\mb_i=(m_{i1}, \dots, m_{iq})$ denotes a vector of
recorded molecular aberrations and $\xb_i=(\mb_i, c_i)$.  
We characterize a subpopulation as a set of mutation-tumor pairs
$\db = \{a: a=(j_a, c_a)$\}, with $j_a \in
\{1,\ldots,q\}$ identifying a molecular aberration and $c_a\in \{1,
\ldots, n_c\}$ denoting tumor type.  
 Each $a \in \db$ denotes a subgroup of patients
with aberration $m_{j_a}=1$ and tumor $c_i=c_a$.  
For example, consider the subpopulation report consisting of
 two subgroups, including  patients with
lung cancer and  BRAF mutation, and patients with breast cancer and
PIK3CA mutation. We denote this subpopulation by \{(BRAF, Lung), (PIK3CA,
Breast)\}. 
 If needed this characterization could incorporate other
baseline covariates 
of interest and/or higher order interactions of covariates to
describe subpopulations. 
However, we do not use such extensions in IMPACT II, restricting 
subsets to be characterized by mutation-tumor pairs $(j_a,c_a)$. 
We add two special cases of subpopulation reports:
let $\db=A_0$ denote the report of no subpopulations {\em and} no overall
treatment effect; and let $\db=A_1$ denote the report of 
an overall treatment effect, but no subpopulations. 
 Note that $A_0$ and $A_1$ are actions, not assumptions about
model parameters. In particular, it is possible that one might report
$A_0$ even when there is a statistically significant treatment effect,
but it is too small to be of clinical relevance. 

 We use a decision
theoretic approach to find an optimal decision $\As$. 
We start by
quantifying the relative preferences across possible reports $\As$. 
That is, we specify
a utility function.  Let $\bth$ denote the parameters of the underlying
survival regression. We still do not need to make any specific
assumptions about the model, except for the existence of such a model.
Given parameter vector $\bth$ and covariate vector $\xb$, we assume
that there is a sampling model $p(y \mid z, \xb,\bth)$ for PFS.  A
utility function is a function $u(\db,\bth)$ for an assumed action $A$ and a
hypothetical true parameter vector $\bth$. In general the utility
function could also depend on observed or future data, but dependence
on $(\db,\bth)$ suffices for the upcoming application.

We define a utility function based on the notion that a clinically
important subpopulation should show a significant treatment effect for
a large population.  Since the response is survival time, it
 is natural   to measure the beneficial treatment effect by the hazard
ratio of TT relative to \Oth.  A minor complication arises from the
fact that in the particular probability model that we shall use a
natural parametrization of hazard ratios does not exist.  In general
hazard ratios depend on time. 
We define a hazard ratio as follows.
Denote by $S(t, z, \xb, \bth)=p(y
\geq t \mid z,\xb,\bth)$ the survival function at time $t$ for a
patient with covariate vector $\xb$ under an assumed model with
parameter vector $\bth$. Similarly, $H(t,z,\xb,\bth)= -\log\{
S(t,z,\xb,\bth) \}$ defines cumulative hazard. Next, we define an
average hazard
\begin{equation}
  \AH(z,\xb,\bth) = H(T, z, \xb, \bth)/T
\label{eq:AH}
\end{equation}
for a chosen horizon $T$.  In our implementation we fix $T$ as the
third quantile of the empirical distribution of the observed PFS times
in the data reported in IMPACT \citep{tsimberidou:12}. 
Using the empirical
distribution of $\xb_i$, we define an $\AH$ for a
mutation-tumor pair $a$ as $\AH_a(a,z,\bth) = \frac{1}{n(a)}
\sum_{i=1}^{n(a)} \AH(z, \xb_i, \bth)$
 (no problem with zero division arises since we will only use AH
for $a$ with $n(a)>0$).   
Here the sum goes over all patients whose covariates fall within the
mutation-tumor pair described by $a$
 (which typically fixes one or several coordinates of $\xb_i$) 
and $n(a)$ denotes the number of
such patients.  
Denote a pair-specific hazard ratio
$$
   \HR(a,\bth) = \frac{\AH_a(a,\Oth,\bth)}{\AH_a(a,TT,\bth)}.
$$
$\HR(a,\bth)=1$ means no difference of average hazard exists between
\Oth\, and TT for mutation-tumor pair $a$.
Under a Cox proportional hazard model for event times $\log
\HR(a,\bth)$ would reduce to just the coefficient for treatment. 
Finally, we assume that the model includes special cases for no
treatment effect for any patient ($H_0$) and for the same treatment
effect for all patients in the eligible population ($H_1$).
Here $H_0$ and $H_1$ are subsets of the parameter space.
Formally, $H_0 = \{\th: \HR(a,\bth)=1 \mbox{ for all } a\}$ and
$H_1 = \{\th: \HR(a,\bth)=c>1 \mbox{ for all } a\}$. One could relax
the definition to allow for approximately equal to $1$ and
approximately constant treatment effect, respectively. 
Importantly, the upcoming discussion does not require prior point
masses, i.e., positive prior probability for $H_0$ or $H_1$.
The choice of $A$ is not directly linked to posterior probabilities in
the sampling model. It is possible that one might want to report
$A_0$, i.e., no effect for any patient, even when $H_0$ is almost surely
not true. This could happen, for example, if we find high posterior
probability for a positive treatment effect, but the effect is
clinically meaningless; or when we find a moderate treatment effect,
but for a very small subpopulation that is impractical for any further
drug development.
Next we will introduce the notion of utilities as a way to formalize
such relative preferences.

We define a utility function,
\begin{equation}
  u(A, \bth) =  
  \begin{cases}
    u_0 I(A=A_0) & \mbox{ if $\th \in H_0$}\\
    \sum_{a\in A}\Big\{ \{ \log[ \HR(a,\bth)
    ]-\beta\}f_{\alpha}(a)\Big\} & \mbox{ otherwise }\\
    u_1 I(A=A_1) & \mbox{ if $\th \in H_1$},
  \end{cases}
  \label{eq:as1}
\end{equation}
where $f_{\alpha}(a)$ is a function on mutation-tumor pairs that
penalizes for  small subgroups   
and $\beta>0$ is a fixed threshold of minimum clinically meaningful
difference in log hazard. 
The broad idea of $u(\db,\bth)$ is to favor the report of large
subgroups with a meaningful treatment effect, and a preference for reporting
the overall null or alternative if appropriate.
However, the specific formalization in \eqref{eq:as1} remains
arbitrary. For example, one could argue to replace 
$\log[\HR(a,\bth)]$ by  $\log[\HR(a,\bth)]-\log[\HR(H_1,\bth)]$, that
is a treatment effect relative to an overall treatment effect. 
In the current application an overall treatment effect of targeted
therapy across all cancers and across all patients is {\it a priori}
unlikely and can be ignored.
For the same reason we do not include a scaling of 
the payoff $u_1$ by a possible overall treatment effect $\log[\HR(H_1,\bth)]$. 
In other applications such modifications could be useful. 

The definition of $f_{\alpha}(a)$ 
 is based on the following considerations. 
Mutation-tumor pairs with small $n(a)$ should be
 penalized, as they are of less clinical interest and at the same
time inference is subject to substantial predictive uncertainty.  
In summary, we use
$$
 f_{\alpha}(a) = \left\{ 
  \begin{array}{l l}
   0 & \quad n(a) < 5 \\
  n(a)^\alpha
 & \quad n(a)\geq 5.
  \end{array} \right.
$$
 Here $n(a)$ denotes the size of the subgroup that is characterized by
$a$. 

In this specification of the utility function, the constants $(u_0,
u_1, \alpha,\beta)$ are tuning parameters.
A practical implementation of the proposed design should use the
following considerations to fix these tuning parameters.
The payoff $u_0$ should be fixed to achieve a desired type I
error rate. That is, 
the fraction of repeat simulations that do not conclude with reporting 
$A^\star=A_0$ under repeated simulations of hypothetical trial
realizations under a simulation truth in $H_0$.
See the upcoming discussion of frequentist operating characteristics
for more details on setting up such repeat simulation. 
Similarly, $u_1$ should be fixed to achieve, or come close to,
a desired true positive
rate, that is, fraction of repeat simulations that end up concluding
with $A^\star=A_1$ under a simulation truth in $H_1$.
The threshold $\beta$ should be elicited from clinical collaborators.
The power $\alpha$ relates to the relative importance of a large
subpopulation. In the implementation we used $\alpha=1/8$,
corresponding to a weak preference for large subpopulations. Any value
$0 \leq \alpha \leq 1$ is reasonable, with $\alpha=0$ implying no
penalty for small subpopulations and $\alpha=1$ implying linearly
increasing utility for larger subpopulations.
Keep in mind the constraint $n(a)\geq 5$, ruling out excessively
small subpopulations.

Among alternative utility functions or criteria to select
subpopulations used in the related literature 
are weighted power \citep{GrafAl:15} in the context of two or few
subgroups under consideration; expected future patient outcome
\citep{SimonSimon:17} in the context of a binary outcome and parametric
Bayesian inference about the benefit of two competing treatments as a
parametric function of baseline covariates; and
enhanced treatment effect in the reported subpopulation versus the
overall population \citep{foster2011subgroup}.
\cite{GrafAl:15} also include size of the proposed
subpopulation in the utility function (penalizing for large subsets
when the treatment entails a safety risk).
For IMPACT II we judged the proposed utility function \eqref{eq:as1} to
best formalize the intention of the study. 

\section{Sampling Model and Adaptive Allocation}
 \subsection{Expected utility and Bayes rule} 
Recall that we assume that there exists a sampling model 
for the observed data, and that the model is indexed by a parameter
vector $\bth$. 
Let $\yb = (y_1,\ldots,y_n)$ denote all observed outcomes, and let 
$\Xb = \{\xb_i, z_i,\; i=1,\ldots,n\}$ denote the known covariates and
treatment assignments.
We write $p(\yb \mid \bth, \Xb)$ for the assumed sampling model.
Now add one more assumption by completing the probability model with a
prior $p(\bth)$ for the unknown parameters, implying a posterior
probability model $p(\bth \mid \yb,\Xb)$.
The utility function $u(\db,\bth)$, together with $p(\bth \mid \yb,\Xb)$
determine the optimal report for a subpopulation as
\begin{equation}
  \As = \arg\max_{\db} \int u(\db,\bth)\; p(\bth \mid \yb,\Xb)\, d\bth.
\label{eq:as}
\end{equation}
In words, the solution $\As$ is the subpopulation report that maximizes the
decision criterion $u(\db,\bth)$. Since $\bth$ is unknown we average
with respect to $p(\bth \mid \yb,\Xb)$. One can argue from first
principles that this is how a rational decision maker should act
\citep{robert:94}. The expectation 
$U(\db) \equiv \int u(\db,\bth)\; p(\bth \mid \yb,\Xb)\, d\bth$, after
integrating out all unknown quantities, is known as expected utility,
and the rule $\As$ is known as the Bayes rule.

\subsection{Nonparametric Bayesian survival regression}
In our implementation we define a nonparametric Bayesian survival regression 
using a model proposed in
\cite{mueller&quintana&rosner:11} and \cite{ana:14}.
A similar model is proposed in
\cite{hannah&blei&powell:11}.
The model is based on a random partition of the experimental units
$[n]=\{1,\ldots,n\}$, in our case, the patients in the study.
That is, patients are arranged in clusters
based on patient-specific covariates $\xb_i$. 
To avoid misunderstanding we note that this partition is unrelated to
the population finding. Any alternative model, without clustering,
could be used. 
We briefly summarize the model below. For more details see 
\cite{mueller&quintana&rosner:11,ana:14}.

Let $S_j \subset [n]$ denote the $j$-th cluster, $j=1,\ldots,J$, and let $\bths_j$
denote cluster-specific parameters.
A cluster-specific sampling model specifies
$p(y_i \mid i \in S_j,\bths_j)$.
Let $\LN(y;\;\mu,\sig^2)$ indicate a lognormal distributed random
variable $y$, that is, $\log(y) \sim \N(\mu,\sig^2)$. In our
implementation we use
$p(y_i \mid i \in S_j,\bths_j=(\mu_j,\sig^2_j)) =
\LN(y_i;\; \mu_j,\sig^2_j)$.
 Censored event times $y_i$ do not introduce any additional
difficulty in posterior inference. Inference will be implemented by
Markov chain Monte Carlo (MCMC) posterior simulation, which allows to
easily accommodate censoring by imputing the missing event
times. Averaging over repeat imputation correctly marginalizes with
respect to $y_i$. 


We complete the model description with a specification for the random partition
$p(S_1,\ldots,S_J \mid \Xb)$ and eventually a prior for $\bths_j$.
Let $\rho_n=\{S_1,\ldots,S_J\}$ denote the random partition (including
the unknown size $J$).
 We use
 \begin{equation}
    p(\rho_n \ \mid \Xb) \propto \prod _{j=1}^{J} g(\xbs_j)\, c(S_j), 
 \label{eq:ppmx}
 \end{equation}
where $\xbs_j= \{\xb_i,\; i \in S_j\}$.
This is a modification of the product partition model (PPM)
$p(\rho_n) \propto \prod c(S_j)$ of \cite{Hart:part:1990}.
With $c(S_j) = M (|S_j|-1)!$ the PPM reduces to the popular Polya urn
model \citep{quintana&iglesias:03}. In \eqref{eq:ppmx} we modified the model to include a factor
$g(\xbs_j)$, which is chosen to favor clusters with similar $\xb_i$.
We refer to $g(\xbs_j)$ as similarity function.
Let $R_i = \{\ell:\; x_{i\ell}\not=\NA\}$
denote the set of recorded covariates for patient $i$ and let
$\xbs_{j\ell} = \{x_{i\ell},\; i \in S_j \mbox{ and } \ell \in R_i\}$.
Let $g_\ell(\xbs_{j\ell})$ denote a similarity function for the
$\ell$-th covariate. We then define
$
g(\xbs_j) = \prod_{\ell=1}^p g_\ell(\xbs_{j\ell}).
$
That is, we define a product similarity function.
For example, if $x_i$ were a single categorical covariate, say tumor
type, letting $M_j$ denote the number of unique values $x_i$ for $i
\in S_j$ and $g(\xbs_j) = 1/M_j$ would define a similarity function
that favors homogeneous clusters with a single unique value $x_i$ in
each cluster.
 See the appendix for the similarity functions that we used in our
implementation, and see 
 \cite{mueller&quintana&rosner:11} for more discussion.
 The described model implements a nonparametric Bayesian
survival regression. Using latent clusters the model allows to
represent arbitrary interactions of covariates and treatment
indicators. However, any alternative survival regression that allows
to learn about enhanced treatment effects, that is, that includes
treatment by covariate interactions, could be used.
For example, one could use the BART survival regression
\citep{BART:16} for which a computationally highly efficient
implementation is available as an R package.

\subsection{Adaptive allocation}

Adaptive treatment allocation in \eqref{eq:alloc} requires the
evaluation of $\pi_i$ as the posterior probability of superiority of
TT over \Oth\, for patient $i$ under the assumed probability model.
Let $y^0_i$ and $y^1_i$ denote potential outcomes for patient $i$ if
the patient were allocated to \Oth\, or TT, respectively and let
$\yb=(y_1,\ldots,y_{i-1})$ and $\Xb=\{\xb_h,z_h,\; h< i\}$ 
denote the data on the first $i-1$ patients.
Then the predictive distribution
\begin{equation}
  p(y^0_{i}, y^1_i \mid \xb_{i}, \Xb,\yb) = 
  \int   p(y^1_i \mid \xb_i, z_i=TT, \bth)\,
    p(y^0_i \mid \xb_i, z_i=\Oth, \bth)\;
  p(\bth \mid \Xb,\yb)\,  
 d\bth,
\label{eq:pred}
\end{equation}
takes the form of an expectation with respect to the
posterior distribution.
See the appendix for more details on \eqref{eq:pred}.
The posterior probability of superiority then becomes  
$$
 \pi_i = \int_{y^0_i < y^1_i} 
   dp(y^0_{i}, y^1_i \mid \xb_{i}, \Xb,\yb).
$$
The attraction of this definition of $\pi_i$ is the evaluation with an
available Monte Carlo sample, without the need for any additional
simulation. This makes it suitable for fast on-line evaluation, as
will be needed for an implementation of the proposed design when a
clinical team has to rely on prompt and uncomplicated evaluation of
allocation probabilities.

Finally, we note that subpopulation finding and adaptive allocation
are two separate features of the proposed design.
One could carry out subpopulation finding alone, without adaptive
allocation and vice versa.
See, for example, \cite{WathenThall:17} or also
\cite{Chappell:03} for recent discussions of the limitations of
response-adaptive designs.
In a large simulation study \cite{WathenThall:17} find only little
evidence for desirable properties of adaptive allocation methods.
Our results are in line with these observations (see Table \ref{simt},
below).

\section{Simulation and Operating Characteristics}
\label{sec:sim}
\subsection{Simulation setup}
We carry out extensive simulation studies to evaluate the model and
the subpopulation finding. We include 6 scenarios specifying different
lognormal regressions with possible interactions among treatment,
mutations, and tumor types.  Using a lognormal regression the
simulation truth is deliberately selected to be a different model than
the assumed PPMx  analysis model.
We use 400 hypothetical patients. 
Table \ref{tab:size} shows the assumed sample sizes, which are chosen
to match the order of magnitude of estimates with the observational
data in IMPACT trial.
For each patient in the simulation we first generate
a treatment indicator $z_i$ with $p(z_i=1) = 0.5$ ($z_i=1$ for TT and
$z_i=0$ for control).
The response $y_i$ is then generated from a lognormal regression model.
Table \ref{tab:scenarios} shows the assumed interaction effects for
each of the six scenarios. Let $p$  $(p=1, 2$ or 3 depending on the
scenario) denote the number of interaction effects for a given 
scenario, let $\beta_j$ denote the corresponding regression coefficient, and let $mc_{ij}\in \{0, 1\}$ denote an indicator whether patient $i$ presents with the combination of mutation and tumor type for the $j$-th interaction. For
example, under scenario 4, $\beta_1 = 0.3$ and $mc_{i1} = 1$ for a patient with PIK3CA mutation and breast cancer. We generate 
$\log(y_i)\sim \mathrm{N}(\beta_0z_i+\sum_{j=1}^p\beta_{j}z_imc_{ij},\sigma^2)$,
where $\sigma=0.2$. 
For each scenario, we simulate 500 trials.

 \begin{table}[!h]
 \centering
 \begin{tabular}{cccc}
  & BRCA     &  Ovary & Lung \\
 \hline
FGFR & 15 & 20 & 5  \\
 \hline
BRAF & 10 &  100 &60  \\
 \hline
 PIK3CA & 50 & 30 &5 \\
  \hline
 PTEN & 13 & 25&5 \\
  \hline
MET & 12& 30&20   \\
 \hline
 \end{tabular}
 \caption{Sample sizes in mutation-tumor pairs. }
 \label{tab:size}
 \end{table}
 
 \begin{table}[!h]
\centering
{\small
\begin{tabular}{ccc}
Scenario & Overall trt     & Interactions  \\
\hline
1 & 0 & none    \\
\hline
2 & 0.4 &   none   \\
\hline
3 & 0 &  BRAF*Lung*z (0.4)  \\
\hline
4 & 0 & PIK3CA*BRCA*z (0.3), BRAF*Lung*z (0.3)\\
&& PTEN*Lung*z(0.4) \\
\hline
5 & 0 & PIK3CA*BRCA*z (0.3), BRAF*Ovary*z (0.4)\\
&&  BRAF*Lung*z(0.3) \\
\hline
 6 & 0 & BRAF*BRCA(0.4), BRAF*Ovary*z (0.3),\\
&& BRAF*Lung*z(0.4)\\
\hline
\end{tabular}
}
\caption{Simulation truth for 6 scenarios. The 2nd column reports the
  true overall treatment effect, that is the regression coefficient $\beta_0$ of
  the treatment indicator $z_i$ in a lognormal regression model.
  The 3rd column report interactions between tumor types, mutations and $z_i$
  (if present). The
values in parentheses are the corresponding regression coefficients $\beta_j$ $(j=1, 2, \mathrm{or} \ 3$, for up to 3 interaction
effects) in the simulation truth.}
\label{tab:scenarios}
\end{table}

We evaluate the proposed inference with respect to two decisions, the
treatment allocation based on \eqref{eq:alloc} and the subpopulation
reports \eqref{eq:as}.
For both evaluations we report summaries under repeated simulations. That
is, we assume a setup where the proposed design is used repeatedly for
multiple trials and performance is evaluated over these repeated
simulations.
Such summaries are known as (frequentist) operating characteristics. 
We use them to calibrate tuning parameters in the utility function. 

\subsection{Adaptive allocation}
\begin{figure}[!ht]
\centering
\includegraphics{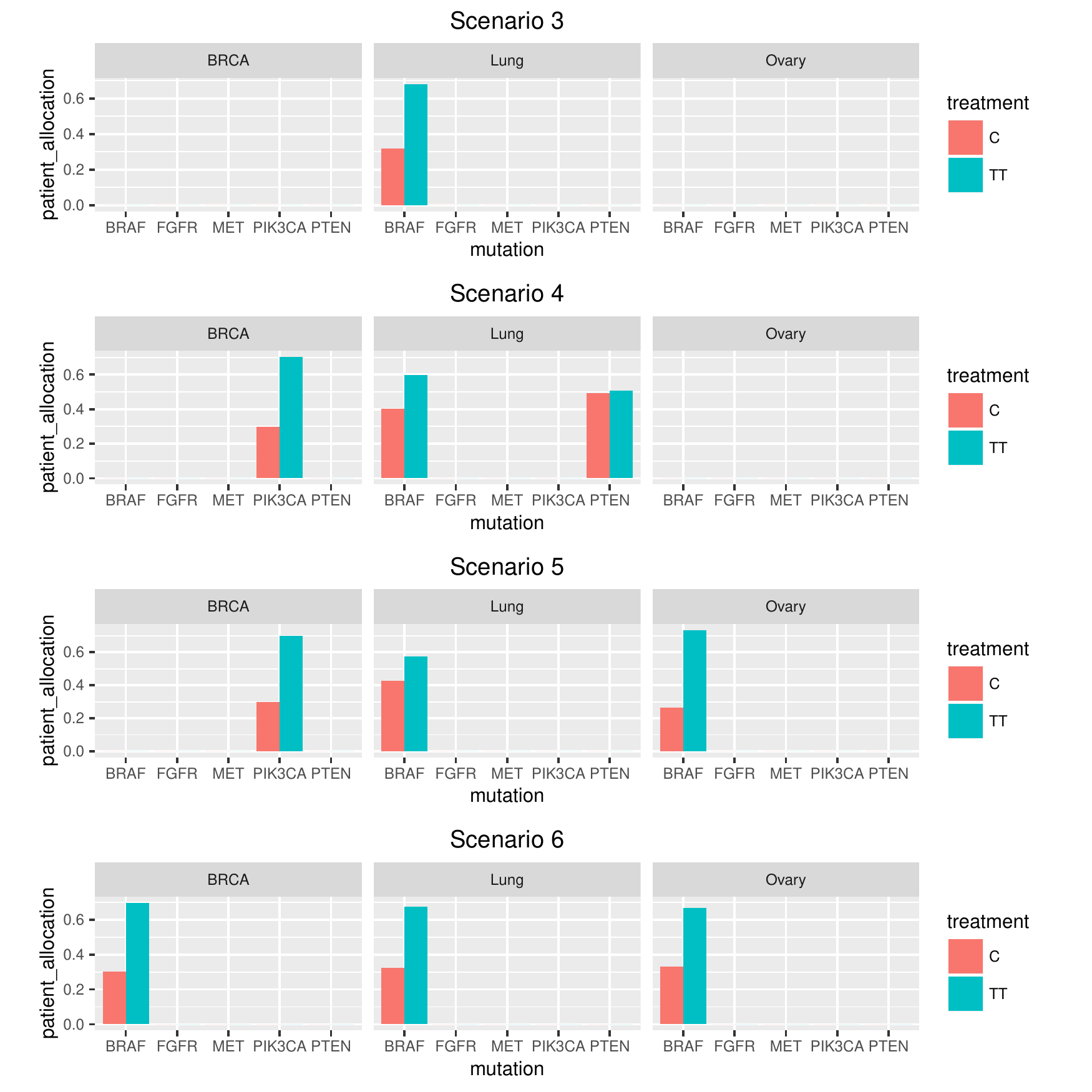}\\
\caption{The average percentage of patients allocated to $\Oth$ and $TT$
  in each mutation-tumor pair with treatment effect different from the overall
  population under the simulation truth.} 
\label{fig:patallo}
\end{figure}

Figure \ref{fig:patallo} plots the average percentage of patients
randomized to TT and \Oth\, for each mutation-tumor pair with
 a corresponding treatment effect that is 
different from the overall population under the simulation truth in
scenarios 3-6. In \underline{scenario 3}, 
where the pair (BRAF, Lung) has a favorable treatment
effect, 68\% of Lung patients with BRAF mutation are randomized to
TT, indicating that more patients receive their superior treatment.
In \underline{scenario 4}, \{(PIK3CA, BRCA), (BRAF, Lung), (PTEN,
Lung)\}  are mutation-tumor pairs with significantly higher treatment
effects. 
For (PIK3CA, BRCA) and (BRAF, Lung),  
70\% and 60\% of all patients 
with these mutation-tumor pairs 
are randomized to TT, respectively. But for lung cancer patients with
PTEN mutation, only 51\% are assigned to TT. The reason is that only
$n(a)=5$ patients have mutation and tumor matching  (PTEN,
Lung), as  shown in Table \ref{tab:size}. The small sample size makes
it difficult to learn about the true effect. 
%
In \underline{scenario 5},  we assume increased treatment effects for
mutation-tumor pairs \{(PIK3CA, BRCA), (BRAF, Ovary), (BRAF, Lung)\}.
All three  pairs include more
than 50 patients. Figure \ref{fig:patallo} shows that more patients in
these groups are allocated to their superior treatments.
Similarly for \underline{scenario 6}.  
Figures S1 through S4 (in the supplementary file) show allocation
probabilities for all mutation-tumor pairs.
 In cases when the simulation truth assumes no differential treatment
effect the  allocation probabilities are close to 0.5.
When additionally the corresponding sample size is large,
e.g., for (BRAF,  Ovary), the allocation probabilities to TT
under repeat simulations are narrowly centered at 0.5 (Figure S1). 

\subsection{Subpopulation finding}
Next we evaluate  rule \eqref{eq:as} for 
subpopulation finding   by summarizing,
again over repeated simulations under a hypothetical truth,  
using in turn each of the 6 scenarios. 
We record how accurately the proposed approach reports true
subpopulations. We have to first define what we understand under a
true subpopulation. 
Let $\bth^0$ denote the true sampling model under one of the 6 scenarios.
We first compute the utility function $U_0(\db) \equiv u(\db,\bth^0)$ for
all possible subpopulations $\db$ under the true sampling model.  
That is, we replace the expectation in \eqref{eq:as} by an
expectation under the true sampling model.
Note that under the fixed hypothetical truth $\bth^0$ there is no
uncertainty left on $\th$. Therefore $U_0$ does
in contrast to $U(\cdot)$ not involve any averaging over $\bth$. 
We then
define the ``true" subpopulation $\dbtrue$ as the top
subpopulation  report with the largest utility $U_0(\db)$. 
The true subpopulation $\dbtrue$ need not match any of the interactions in
the simulation truth in Table \ref{tab:scenarios}.

This separation of the statistical inference related to model fit and
estimation versus the decision is important. It is related to the
difference between statistical significance versus practical
relevance, but goes beyond that. For the model fit we use a maximally
flexible model that should ideally be able to fit higher order
interactions and more.
In contrast, for the subgroup report we prefer a simple and
parsimonious solution. This preference is formalized by the utility
function. 

\begin{figure}[!h]
  \centering
    \includegraphics[scale=0.9]{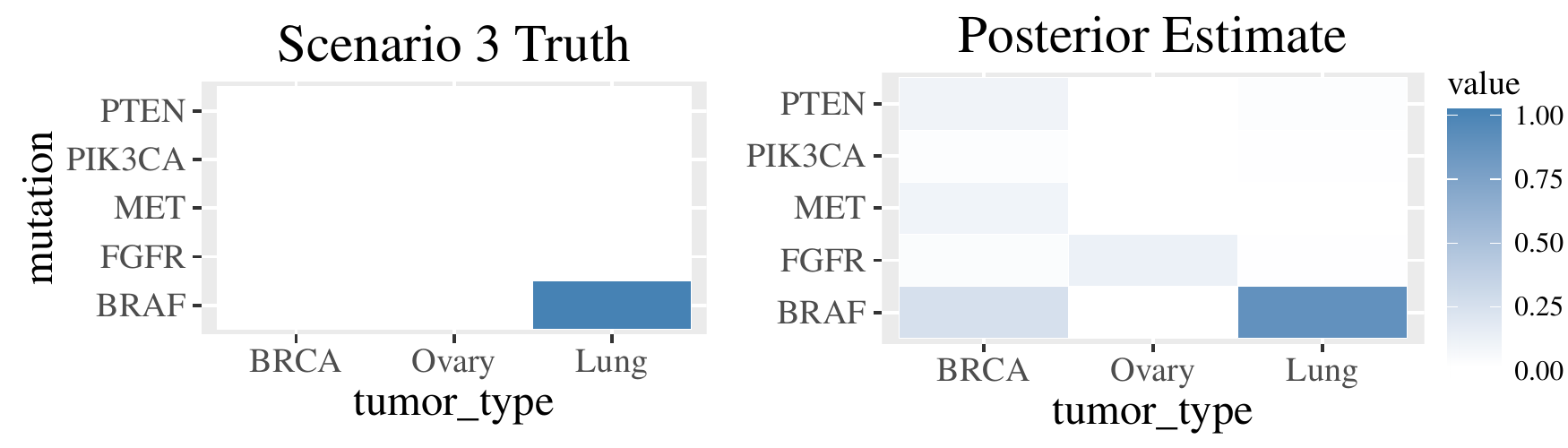}\\
    \includegraphics[scale=0.9]{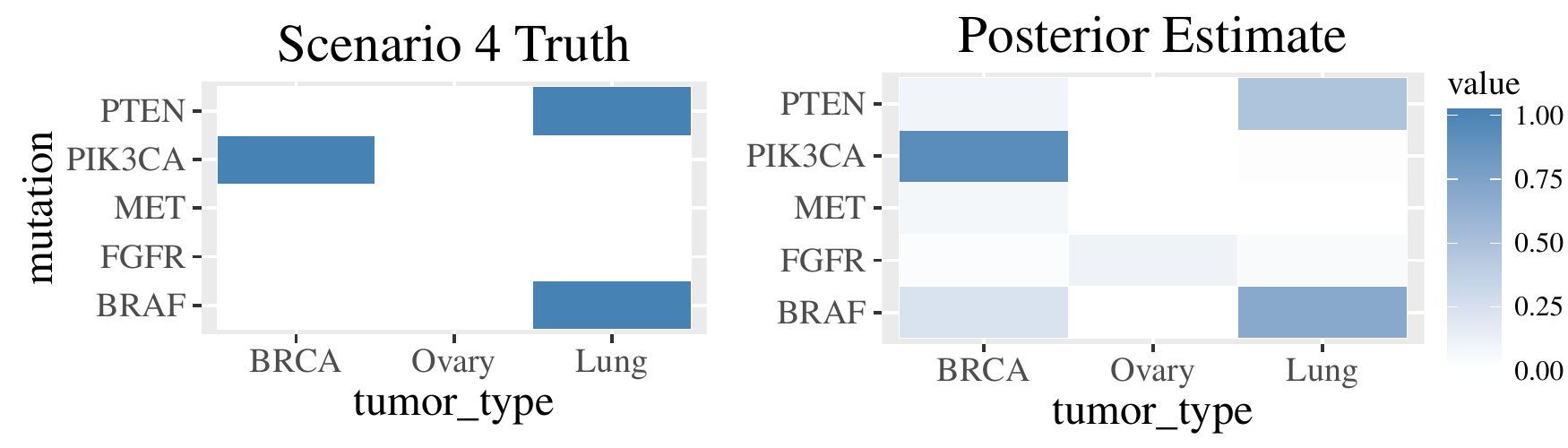}\\
    \includegraphics[scale=0.9]{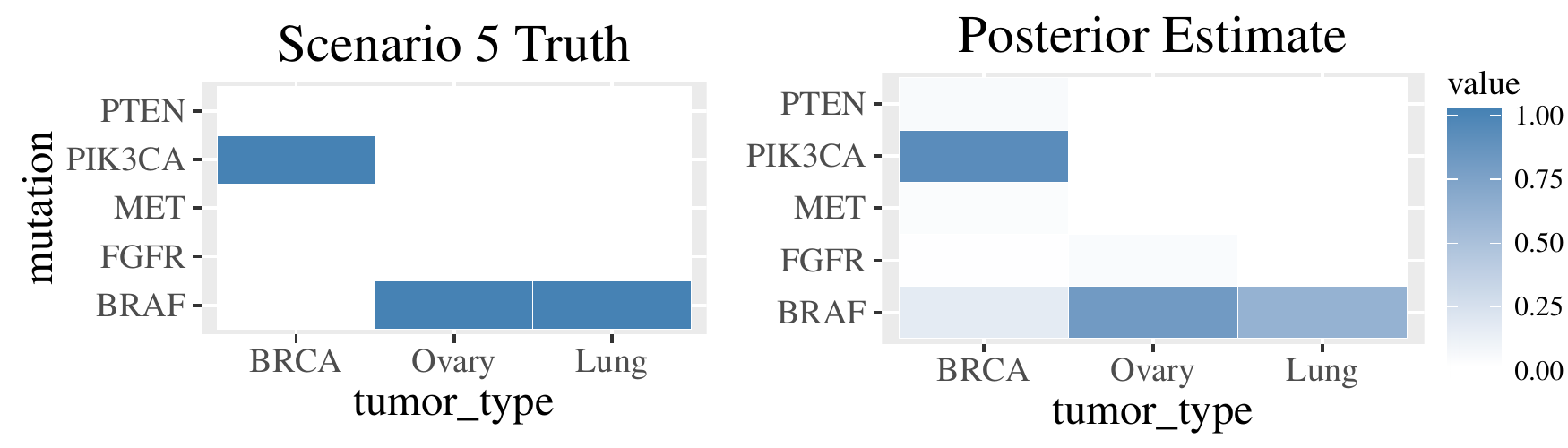}\\
    \includegraphics[scale=0.9]{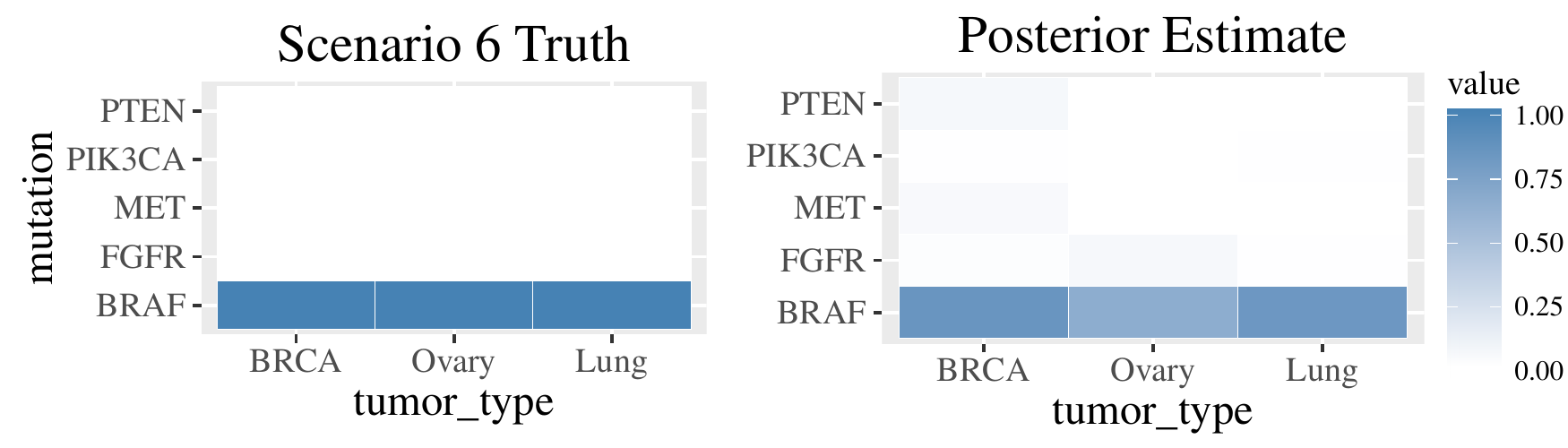}\\
  \caption{In each scenario, the left panel shows the simulation
    truth. Blue cells represent mutation-tumor pairs with treatment
    effect different from the overall population under the simulation truth;
    the right panel shows a heatmap of $Pr(a)$, the probability (under
    repeated simulation) of reporting each mutation-tumor pair.}
  \label{fig:popufind}
\end{figure}

In each scenario, we compute the percentage of trials in which each
subgroup $a$ is reported: 
$$
  Pr(a) = \frac{1}{500}\sum_h I(a\in \As_h).  
$$
Here $\As_h$ is the report in repeat simulation $h$, 
$h=1,\ldots,500$, using the Bayes rule \eqref{eq:as}. 
In words, $Pr(a)$ are estimated (frequentist) probabilities over
repeat simulations and $h$ indexes each simulation. 
Figure \ref{fig:popufind} shows
$Pr(a)$, that is $\As$ (right panel in each pair of panels), versus the simulation truth
$U_0(\db)$ (left panel). 

\subsection{Operating characteristics}
The utility function depends on the parameters $u_0, u_1$, $\alpha$,
and $\beta$.  We fix these parameters to achieve a desired error rate.  
For this purpose,  we summarize several
types of error rates.  Recall that $A_0$ indicates the decision not to
report any recommended subpopulation and $A_1$ indicates the decision
to report the entire patient population.
And recall the notation $\db = \{a: a=(j_a, c_a)$\} for any other subpopulation report.
We will use superscript $^c$ to denote the absence of a
particular report in the list  of pairs in each subpopulation report.
Finally, as before we use $\Pr$ to denote a frequentist rate of the
various errors. That is, the probability under repeated simulations.
 And we slightly abuse the notation of conditioning bar. In
$\Pr(a \mid b)$, the first argument, $a$ refers to a decision, and the
second, $b$, refers to an event under the simulation truth. 
For example 
$\Pr(a^c \mid a)= \sum_hI(a\notin \As_h)/500$ 
refers to the probability of not reporting mutation-tumor pair $a$
($a\notin \As_h$) when the pair is in the true subpopulation, i.e., $a\in
\dbtrue$.  The probability is evaluated as average over 500
repeated trials. 
When $\dbtrue$ includes multiple mutation-tumor pairs the rates
include an average over all $a \in \dbtrue$, as indicated below. 

We report the following six error rates:
1) Type 1 error (TIE) $ = \Pr(A_0^c \mid H_0)$; 
2) TSR (true subgroup rate)  $ = \Pr(a \mid a) = \sum_{a\in \dbtrue}\sum_hI(a\in \As_h)/(500\times|\dbtrue|);      $  
3) TPR (true positive rate)  $ = \Pr(A_1 \mid  H_1);$ 
4) FSR (false subgroup rate) $ = \Pr(a \mid a^c)= \sum_{a\notin \dbtrue}\sum_hI(a\in \As_h)/(500\times|(\dbtrue)^c|);    $
5) FNR (false negative rate) $ = \Pr( A_0 \mid H_0^c); $ 
6) FPR (false positive rate) $ = \Pr(A_1 \mid H_1^c).   $ 
 The selection of these error rates could be changed as
desired. For example, for other applications it might be meaningful to
report expected (under repeated simulation) false discovery rates, etc. 
All unknown parameters in the utility function (\ref{eq:as1}) are calibrated to restrict TIE=0.05, shown in scenario 1 and TPR = 0.9, shown in scenario 2, $u_0=1.3, u_1=20, \alpha=1/8$, and $\beta=0.4$. 
Note that TIE and TPR are special cases of TSR, and FNR and FPR
are special cases of FSR.
Not all error rates are meaningful in all scenarios. For example, TIE
is only meaningful when $H_0$ is in fact the true population
under $U_0(\db)$ and similar for TPR.
Table \ref{simt} summarizes the 6 error rates in the 6 scenarios.

\begin{table}[ht]
  \begin{center}
    \begin{tabular}{cccccccc}
      \hline
      & Scenario &   TIE  & TSR & TPR &  FSR & FNR  & FPR\\
       \hline
      & 1 & 0.05   & -    & - &- &- &- \\
       \hline
      & 2 & -  & -    &.90&-&- &- \\
       \hline
      & 3 & -  & .87    &- &.04&.10 &.00 \\
       \hline
      & 4 & -    & .68    & - &.04 &.04 &.00\\
       \hline
      & 5 & -  & .77    &- &.02&.01 &.00 \\
       \hline
      &6 & -  & .77    &- &.02&.04 &.00 \\
       \hline
       &3 (without AR) & -  & .86    &- &.01&.14 &.00 \\
       \hline
    \end{tabular}
  \end{center}
  \caption{Simulation operating characteristics results.
    The table shows six different error rates for the subpopulation
    finding report, including
    TIE  $ = \Pr(A_0^c \mid H_0)$; 
    TSR  $ = \Pr(a   \mid a); $
    TPR  $ = \Pr(A_1 \mid  H_1);$ 
     FSR  $ = \Pr(a   \mid a^c);    $
    FNR  $ = \Pr(A_0 \mid H_0^c); $  and  
    FPR  $ = \Pr(A_1 \mid H_1^c).   $
    AR denotes adaptive randomization. 
 }
  \label{simt}
\end{table}

Finally, we investigate the effect of the adaptive randomization
(AR). We consider scenario 3 in the simulation study, but now without
AR.  The results are reported as an additional line in Table
\ref{simt}.
Compared to the simulation with AR the changes are small. This is
probably due to the fact that adaptive randomization is conservative
(bounded by $p_0$ and $p_1$, respectively), the sample size is
moderate in each subpopulation, and the model includes borrowing of
strength across different subpopulations.


\subsection{Inference and Comparison} 

For comparison, we implement two alternative trial designs: 
a simple two-arm randomization (NAIVE) and separate trials for each
molecular aberration (SEPARATE).
NAIVE assigns patients equally to TT and \Oth, and compares TT
with \Oth\, over the whole population. In particular, no subgroups are
considered in the NAIVE design. 
In the NAIVE design, we assume $\log(y_i)\mid z_i\sim N(\mu_{z_i}, \sigma^2_{z_i})$ with conjugate priors $\mu_{z_i}\sim N(\mu_0, \tau^2)$ and $\sigma^2_{z_i}\sim \mathrm{Inverse \ Gamma}(b_1, b_2)$, where $z_i$ = TT or O. 
In the SEPARATE design, we perform
separate independent studies for each subgroup determined by mutation
only, that is, ``rows" in Figure 3. In other words, SEPARATE are five
separate trials with the NAIVE design. 

We compare the three methods based on
the expected PFS time of a hypothetical future patient who is
assigned the optimal treatment as estimated from these three methods.  
Subtracting (true) expected PFS under \Oth, any comparison of
expected PFS under the optimal treatment under different designs can
equivalently be interpreted as a difference
in treatment effects, defined as difference under optimal treatment and
\Oth\, (thus the acronym TE, below). 
Let $E_y$ denote an expectation with respect to $y$ under the
simulation truth, and then define
 $\TE(\xb,z) = E_y(y\mid \xb, z)$ 
to be the expected treatment effect for a future patient with
covariate $\xb$ under treatment $z$. 
Let $\TE_{\xb} = \TE(\xb, z^*_{\xb})$ be $\TE$ under the optimal treatment 
$z^*_{\xb}$ for a patient with covariate $\xb$, as inferred from posterior
inference under the analysis model. 
For instance, the optimal treatment for patients in
the reported subpopulation $A^*$ in \eqref{eq:as} is TT, otherwise
\Oth. 
Similarly let $E_{y\mid \Xb}$ denote an expectation over $y$ with
respect to the posterior predictive distribution (under the analysis
model), and let $\TEhat(\xb,z) = E_{y \mid \Xb} (y \mid \xb, z)$
denote the estimated $\TEhat$ under treatment $z$ and $\TEhat_{\xb} =
\TEhat(\xb,z^*_{\xb})$.
For NAIVE and SEPARATE designs, we compute the optimal treatment for
a patient with covariate $\xb$ as
$z^*_{\xb}=\mathrm{argmax}_z \TEhat(\xb,z)$. 
That means, if 
$\TEhat(\xb,z=\TT)>\TEhat(\xb,z=\mathrm{O})$ then
$z^*_{\xb}=\TT$; otherwise $z^*_{\xb}=\mathrm{O}$.
Finally, we define
$\TE_a=\TE_{\xb}$ for $\xb=(\mb, c)$ with $c=c_a, m_{j_a}=1$, 
and $m_j=0, j\not=j_a$ 
to be the expected  PFS  for the mutation-tumor pair $a=(j_a, c_a)$
under the simulation truth and $\TEhat_{a}$ to be the
estimated PFS.  And, again, subtracting true PFS under \Oth, each of
these summaries can be considered a summary on treatment effects. 

\begin{figure}[!h]
\centering
\includegraphics[width=.8\textwidth]{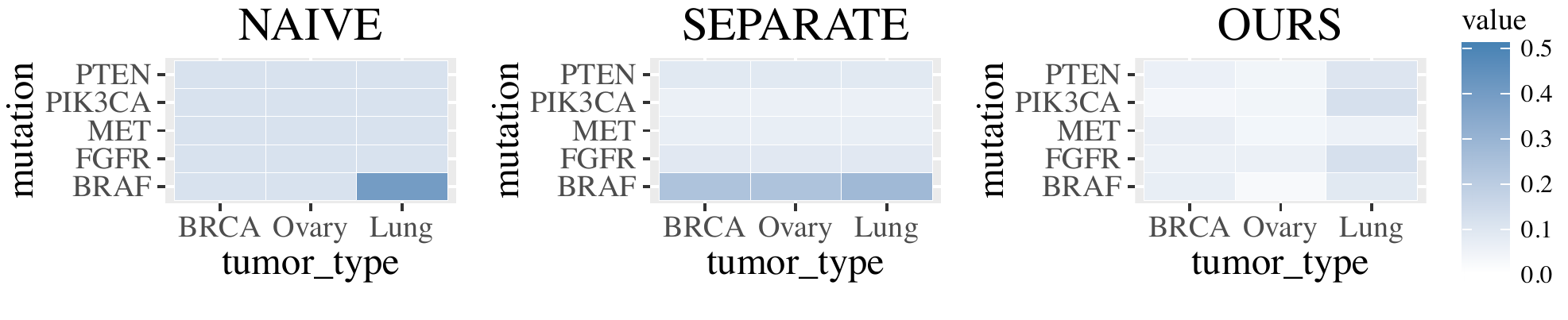} \\
Scenario 3 \\[4pt]
\includegraphics[width=.8\textwidth]{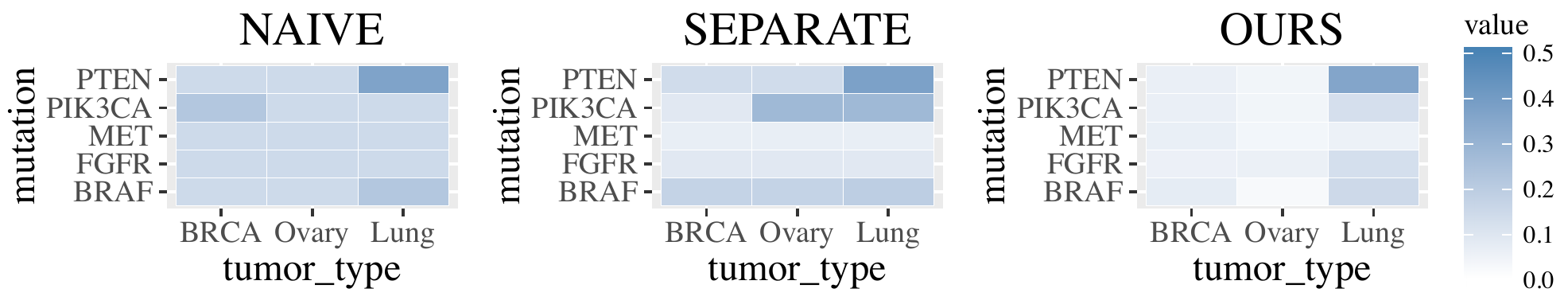}\\
Scenario 4 \\[4pt]
\includegraphics[width=.8\textwidth]{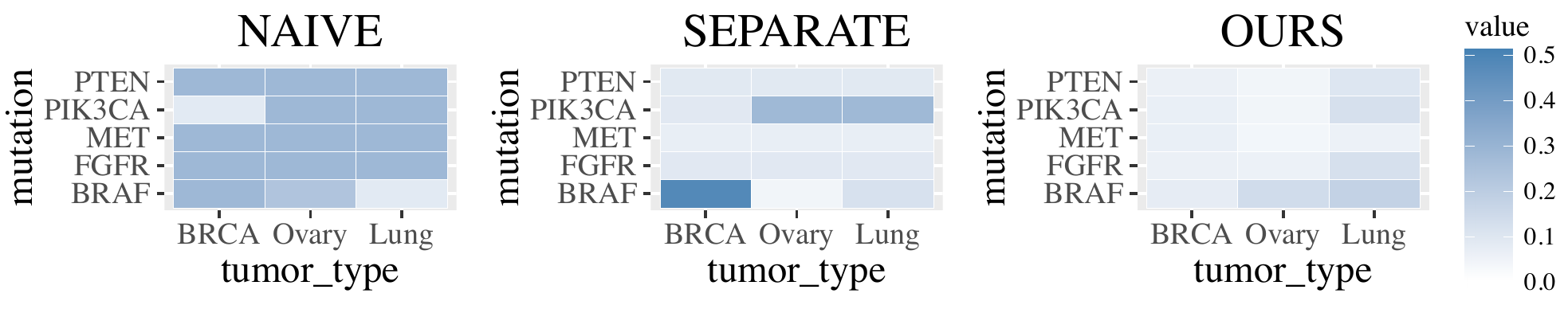}\\
Scenario 5 \\[4pt]
\includegraphics[width=.8\textwidth]{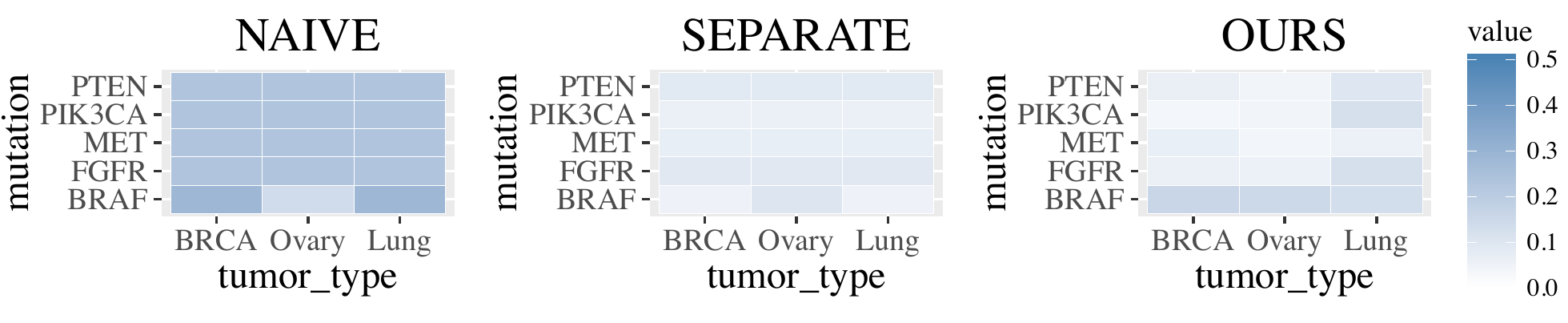}\\
Scenario 6  
\caption{The differences between the estimated treatment effect and true treatment effect for each mutation-tumor pair under NAIVE, SEPARATE, and OURS in scenarios 3-6.} 
\label{fig:compare}
\end{figure}

Figure \ref{fig:compare} 
plots $|\TEhat_{a}-\TE_a|$ for each mutation-tumor
pair under the three methods: NAIVE, SEPARATE, and OURS in scenarios 3-6. 
OURS refers to the proposed approach.  
We find that OURS reports the smallest differences among all
mutation-tumor pairs in scenarios 3-5.  SEPARATE performs slightly
better than OURS in scenario 6 since the simulated true mutation-tumor
pairs (\{(BRAF, BRCA), (BRAF, Ovary), (BRAF, Lung) \}) with treatment
effect different from the overall population happen to match the
analysis model of SEPARATE, which considers the subgroup by mutation
only.  

\section{Conclusion}
\label{sec:conclusion}
We have outlined a Bayesian adaptive clinical trial design to assign patients to their superior treatment and 
a practicable decision theoretic approach to optimal subpopulation finding. 
The strengths of the proposed approach include: 1) we make decisions
based on a flexible utility function that reflects the requirement of
the clinicians appropriately and realistically, such as rewarding the
correct subpopulation reports and penalizing small size
subpopulations; 2) we use a general class of probability models that
efficiently incorporate treatment covariate as well as
covariate-covariate interactions.

Some limitations remain.
For example, the solution of the proposed decision theoretic approach
depends on the often arbitrary choice of the unknown parameters in
utility function. The problem is mitigated by calibrating frequentist
operating characteristics like TIE.

Finally, we note that the proposed approach was introduced in an
oncology trial, but is of course valid in any other disease area.
Similarly, we introduced specific choices for the utility function,
sampling model, and prior. But others could be used, including in
particular, informative priors when available. 
In the nonparametric Bayesian survival regression that we used the 
easiest way to include informative priors is to include hypothetical
additional patients in the data for posterior computation.

\appendix
\section*{ Appendix: The PPMx model}
\paragraph{\bf Similarity function.}
\cite{mueller&quintana&rosner:11} propose a generic construction
of similarity functions in \eqref{eq:ppmx} based on an auxiliary probability model
$q(\xbs_j \mid \xis_j)$ and $q(\xis_j)$.
Here $\xis_j$ are additional parameters used for the definition of the
similarity function.
Importantly, the model $q(\cdot)$ is used only to obtain an easy
expression for the similarity function, without any notion of modeling
a distribution of covariates $\xb_i$.
We define
$ 
  g(\xb^*_j)=\int \prod_{i\in S_j}q(\xb_i\mid \xis_j)q(\xis_j) \,d\xis_j,
$ 
which can be interpreted as the marginal model under $q(\cdot)$, and
can be analytically evaluated when the distributions are chosen as
conjugate pair.
The definition is meaningful when the marginal is highest for sets of
covariate values $\xbs_j$ that would be considered to be similar, as
is the case under most models.

For {\em continuous} $x_i$ we 
define $\xibs_j=(\mu_j,v_j)$. Let $q(x_i\mid 
    \xibs_j) = N(x_i\mid\; \mu_j,v_j)$, and $q(\mu_j,v_j)$ be 
    the conjugate normal-inverse chi-square (or gamma) prior distribution 
    \citep[see, for example,][]{carlin&gelman&stern&rubin}. Then 
    $g(\xb^*_j)$ is a scaled and correlated $n_j$-dimensional 
    multivariate $t$ density. Here $n_j=|S_j|$ is the size of the 
    $j$-th cluster.
Next, consider a {\em categorical} covariate $x_i$ with $c$ 
    levels, $x_i \in \{1,\ldots,c\}$. Let $\xibs_j = 
    (\pi_1,\ldots,\pi_c)$ where $0\le\pi_r$ for all 
    $r=1,\ldots,c$ and $\sum_{r=1}^c\pi_r=1$. Then use $q(x_i\mid 
    \xibs_j)\equiv\mult(x_i\mid 1,\xibs_j)$ and 
    $q(\xibs_j)\equiv\Dir(\xibs_j\mid \balpha)$ for some suitable choice of 
    $\balpha$. In this case, $q(\xibs_j\mid  \xb^*_j)$ is again a 
    Dirichlet distribution, and $g(\xb^*_j)$ is a 
    Dirichlet-multinomial distribution. In the particular binary 
    case ($c=2$) we get the beta-binomial distribution. 
For {\em Count covariates}, we define
    $q(x_i\mid\xis_j)$ as a Poisson distribution with rate
    $\xis>0$, and for $q(\xis)$ we assume a gamma distribution. Then, 
    $q(\xis_j \mid \xb^*_j)$ is again a gamma distribution and 
    $g(\xb^*_j)$ reduces to the Poisson-gamma distribution.  
%

Under all three cases we can exploit the conjugate nature of
$q(\cdot)$ and use Bayes theorem to evaluate
$ 
  g(\xb^*_j)={\prod_{i\in S_j}q(\xb_i\mid \xits_j)q(\xits_j)}\big/
  {q(\xits_j\mid\xbs_j)},
$ 
where $\xits_j$ is {\em any} fixed value of $\xis_j$. Note that 
this expression can be readily evaluated and the dimension of 
$\xits_j$ does not depend on the cluster size. 

\paragraph{\bf Posterior predictive inference.}
In \eqref{eq:pred} we use the posterior predictive distribution under
the assumed model. We briefly describe 
$p(y_{n+1} \mid \xb_{n+1},z_{n+1},\yb,\Xb)$ (recall that $\Xb$
includes the treatment assignments $z_1,\ldots,z_n$) under the PPMx model.
In words, the posterior predictive distribution 
averages with respect to the cluster membership for $i=n+1$,
with respect to the posterior distribution on the
cluster-specific parameters $\bths$ and finally, with respect to the
posterior distribution on the random partition.
The latter average reduces to a sum over all possible partitions of
$[n]$. 

As before, let $\rho_n=\{S_1,\ldots,S_J\}$ denote the random partition
(including the random size $J$ of the partition). 
That is, there are $J$ clusters $S_1, \ldots, S_J$ with 
$\bigcup_{j=1}^J S_j = [n]$. Let $\xbs_j=\{\xb_i,\; i\in S_j\}$
denote the covariates arranged by clusters.
We first match $i=n+1$ with one of the $J$ current clusters based on
matching $\xb_{n+1}$ with $\xbs_j$. Conditional on $n+1 \in S_j$ and
conditional on $\bths$ the prediction for $y_{n+1}$ is 
$p(y_{n+1} \mid n+1 \in S_j, \bths_j) = \LN(\mu_j,\sig^2_j)$.
The desired predictive distribution 
$p(y_{n+1} \mid \xb_{n+1},z_{n+1}, \Xb,\yb)$ is then defined by averaging 
$p(y_{n+1} \mid n+1 \in S_j, \bths_j)$ with respect to 
the cluster-specific parameters and with respect to the random
partition.  In summary,
\begin{multline}
  p(y_{n+1} \mid \xb_{n+1}, z_{n+1}, \Xb,\yb) =
  \sum_{\rho_n}
  p(\rho_n \mid \Xb,\yb)\;
  \int 
  p(\bths \mid \rho_n, \Xb,\yb)\,  \\
  \times
  \left \{ \sum_{j=1}^{J+1} 
    p(y_{n+1} \mid n+1 \in S_j, \bths_j)\;
    p(n+1 \in S_j \mid \xb_{n+1}, z_{n+1}, \Xb,\rho_n)\;
  \right\}\; d\bths.
\nonumber 
\end{multline}
The innermost sum is the average with respect to the cluster
membership for the $(n+1)$-st patient. Note that the cluster
membership includes a regression on $\xb_{n+1}$. We allocate the next
patient with higher probability to existing clusters $S_j$ with
similar covariates $\xbs_j$.
Also note that the average includes $j=J+1$, that is, the possibility
that $(n+1)$ forms a new (singleton) cluster $S_{J+1}=\{n+1\}$.

\paragraph{\bf Software.}
In implementation of the proposed design as R macros can be found at 
\url{http://www.ams.jhu.edu/~yxu70/software.html}.

\begin{acknowledgement}
Peter M\"uller and Yanxun Xu's research is partly supported by NIH grant R01 CA132897.
\end{acknowledgement}
\vspace*{1pc}

\noindent {\bf{Conflict of Interest}}

\noindent {\it{The authors have declared no conflict of interest. }}

\bibliographystyle{apalike}
\bibliography{subgroup}

\begin{thebibliography}{}

\bibitem[Baladandayuthapani et~al., 2010]{baladandayuthapanibayesian2010}
Baladandayuthapani, V., Ji, Y., Talluri, R., {Nieto-Barajas}, L., and Morris,
  J. (2010).
\newblock {B}ayesian random segmentation models to identify shared copy number
  aberrations for array {CGH} data.
\newblock {\em Journal of the American Statistical Association},
  105:1358--1375.

\bibitem[Barker et~al., 2009]{barker2009spy}
Barker, A., Sigman, C., Kelloff, G., Hylton, N., Berry, D., and Esserman, L.
  (2009).
\newblock {I-SPY} 2: an adaptive breast cancer trial design in the setting of
  neoadjuvant chemotherapy.
\newblock {\em Clinical Pharmacology and Therapeutics}, 86(1):97--100.

\bibitem[Barski and Zhao, 2009]{barskigenomic2009}
Barski, A. and Zhao, K. (2009).
\newblock Genomic location analysis by {ChIP-Seq.}
\newblock {\em Journal of Cellular Biochemistry}, 107:11--18.

\bibitem[Berry et~al., 2013]{berryAl13}
Berry, S.~M., Broglio, K.~R., Groshen, S., and Berry, D.~A. (2013).
\newblock {{B}ayesian hierarchical modeling of patient subpopulations:
  efficient designs of {P}hase {I}{I} oncology clinical trials}.
\newblock {\em Clin Trials}, 10(5):720--734.

\bibitem[Brannath et~al., 2009]{BrannathAl:09}
Brannath, W., Zuber, E., Branson, M., Bretz, F., Gallo, P., Posch, M., and
  Racine-Poon, A. (2009).
\newblock Confirmatory adaptive designs with bayesian decision tools for a
  targeted therapy in oncology.
\newblock {\em Statistics in Medicine}, 28(10):1445--1463.

\bibitem[Bretz et~al., 2006]{BretzAl:06}
Bretz, F., Schmidli, H., König, F., Racine, A., and Maurer, W. (2006).
\newblock Confirmatory seamless phase ii/iii clinical trials with hypotheses
  selection at interim: General concepts.
\newblock {\em Biometrical Journal}, 48(4):623--634.

\bibitem[Conley and Doroshow, 2014]{conley2014molecular}
Conley, B.~A. and Doroshow, J.~H. (2014).
\newblock Molecular analysis for therapy choice: Nci match.
\newblock In {\em Seminars in Oncology}, volume~41, pages 297--299. Elsevier.

\bibitem[Curtis et~al., 2012]{curtisthe2012}
Curtis, C., Shah, S., Chin, S., Turashvili, G., Rueda, O., Dunning, M., and
  et~al. (2012).
\newblock The genomic and transcriptomic architecture of 2,000 breast tumours
  reveals novel subgroups.
\newblock {\em Nature}, 486:346--352.

\bibitem[Dixon and Simon, 1991]{dixon1991bayesian}
Dixon, D.~O. and Simon, R. (1991).
\newblock {B}ayesian subset analysis.
\newblock {\em Biometrics}, 47:871--881.

\bibitem[Foster et~al., 2011]{foster2011subgroup}
Foster, J.~C., Taylor, J.~M., and Ruberg, S.~J. (2011).
\newblock Subgroup identification from randomized clinical trial data.
\newblock {\em Statistics in Medicine}, 30(24):2867--2880.

\bibitem[Gelman et~al., 2004]{carlin&gelman&stern&rubin}
Gelman, A., Carlin, J.~B., Stern, H.~S., and Rubin, D.~B. (2004).
\newblock {\em {B}ayesian data analysis}.
\newblock Texts in Statistical Science Series. Chapman \& Hall/CRC, Boca Raton,
  FL, second edition.

\bibitem[Graf et~al., 2015]{GrafAl:15}
Graf, A.~C., Posch, M., and Koenig, F. (2015).
\newblock Adaptive designs for subpopulation analysis optimizing utility
  functions.
\newblock {\em Biometrical Journal}, 57(1):76--89.

\bibitem[Hannah et~al., 2011]{hannah&blei&powell:11}
Hannah, L., Blei, D., and Powell, W. (2011).
\newblock Dirichlet process mixtures of generalized linear models.
\newblock {\em Journal of Machine Learning Research}, 12:1923–1953.

\bibitem[Hartigan, 1990]{Hart:part:1990}
Hartigan, J.~A. (1990).
\newblock Partition models.
\newblock {\em Communications in Statistics: Theory and Methods},
  19:2745--2756.

\bibitem[Hudis, 2007]{hudis2007trastuzumab}
Hudis, C.~A. (2007).
\newblock Trastuzumab mechanism of action and use in clinical practice.
\newblock {\em New England Journal of Medicine}, 357(1):39--51.

\bibitem[Hyman et~al., 2015]{hyman2015vemurafenib}
Hyman, D.~M., Puzanov, I., Subbiah, V., Faris, J.~E., Chau, I., Blay, J.-Y.,
  Wolf, J., Raje, N.~S., Diamond, E.~L., Hollebecque, A., et~al. (2015).
\newblock Vemurafenib in multiple nonmelanoma cancers with braf v600 mutations.
\newblock {\em New England Journal of Medicine}, 373(8):726--736.

\bibitem[Karrison et~al., 2003]{Chappell:03}
Karrison, T.~G., Huo, D., and Chappell, R. (2003).
\newblock A group sequential, response-adaptive design for randomized clinical
  trials.
\newblock {\em Controlled Clinical Trials}, 24(5):506 -- 522.

\bibitem[Misale et~al., 2012]{misale2012emergence}
Misale, S., Yaeger, R., Hobor, S., Scala, E., Janakiraman, M., Liska, D.,
  Valtorta, E., Schiavo, R., Buscarino, M., Siravegna, G., et~al. (2012).
\newblock Emergence of {KRAS} mutations and acquired resistance to anti-{EGFR}
  therapy in colorectal cancer.
\newblock {\em Nature}, 486(7404):532--536.

\bibitem[M{\"u}ller et~al., 2011]{mueller&quintana&rosner:11}
M{\"u}ller, P., Quintana, F.~A., and Rosner, G.~L. (2011).
\newblock {A Product Partition Model with Regression on Covariates}.
\newblock {\em Journal of Computational and Graphical Statistics},
  20(1):260--278.

\bibitem[Quintana and Iglesias, 2003]{quintana&iglesias:03}
Quintana, F.~A. and Iglesias, P.~L. (2003).
\newblock {{B}ayesian Clustering and Product Partition Models}.
\newblock {\em Journal of the Royal Statistical Society Series B}, 65:557--574.

\bibitem[Quintana et~al., 2014]{ana:14}
Quintana, F.~A., M\"uller, P., and Papoila, A.~L. (2014).
\newblock Cluster-specific variable selection for product partition models.
\newblock Technical report, Pontificia Universidad Catolica de Chile.

\bibitem[Robert, 1994]{robert:94}
Robert, C. (1994).
\newblock {\em The {B}ayesian Choice}.
\newblock Springer-Verlag.

\bibitem[Ruberg et~al., 2010]{ruberg2010mean}
Ruberg, S.~J., Chen, L., and Wang, Y. (2010).
\newblock The mean does not mean as much anymore: finding sub-groups for
  tailored therapeutics.
\newblock {\em Clinical Trials}, 7(5):574--583.

\bibitem[Schnell et~al., 2017]{Schnell:17}
Schnell, P., Tang, Q., Müller, P., and Carlin, B.~P. (2017).
\newblock Subgroup inference for multiple treatments and multiple endpoints in
  an alzheimer’s disease treatment trial.
\newblock {\em Ann. Appl. Stat.}, 11:949--966.

\bibitem[Schnell et~al., 2016]{Schnell:16}
Schnell, P.~M., Tang, Q., Offen, W.~W., and Carlin, B.~P. (2016).
\newblock A bayesian credible subgroups approach to identifying patient
  subgroups with positive treatment effects.
\newblock {\em Biometrics}, 72:1026--1036.

\bibitem[Simon and Simon, 2017]{SimonSimon:17}
Simon, N. and Simon, R. (2017).
\newblock Using bayesian modeling in frequentist adaptive enrichment designs.
\newblock {\em Biostatistics}, in press.

\bibitem[Simon, 2002]{simon2002bayesian}
Simon, R. (2002).
\newblock {B}ayesian subset analysis: application to studying
  treatment-by-gender interactions.
\newblock {\em Statistics in Medicine}, 21(19):2909--2916.

\bibitem[Simon, 2012]{simon12}
Simon, R. (2012).
\newblock {{C}linical trials for predictive medicine}.
\newblock {\em Statistics in Medicine}, 31(25):3031--3040.

\bibitem[Sivaganesan et~al., 2011]{sivaganesan2011bayesian}
Sivaganesan, S., Laud, P.~W., and M{\"u}ller, P. (2011).
\newblock A {B}ayesian subgroup analysis with a zero-enriched polya urn scheme.
\newblock {\em Statistics in Medicine}, 30(4):312--323.

\bibitem[Sivaganesan et~al., 2013]{laud2013subgroup}
Sivaganesan, S., Laud, P.~W., and M{\"u}ller, P. (2013).
\newblock Subgroup analysis.
\newblock In Damien, P., Dellaportas, P., Polson, N., and Stephens, D.,
  editors, {\em {B}ayesian Theory and Applications}, pages 576--592. Oxford
  University Press.

\bibitem[Snijders et~al., 1998]{Snij}
Snijders, A., Nowak, N., Segraves, R., Blackwood, S., Brown, N., and et~al.
  (1998).
\newblock {Assembly of microarrays for genome-wide measurement of DNA copy
  number}.
\newblock {\em Nature Genetics}, 29:263--264.

\bibitem[Sparapani et~al., 2016]{BART:16}
Sparapani, R., Logan, B., McCulloch, R., and Laud, P. (2016).
\newblock Nonparametric survival analysis using bayesian additive regression
  trees (bart).
\newblock {\em Statistics in Medicine}, 35(16):2741--2753.

\bibitem[Tsimberidou et~al., 2012]{tsimberidou:12}
Tsimberidou, A., N.G., I., Hong, D., Wheler, J., Falchook, G., Fu, S.,
  Piha-Paul, S., Naing, A., Janku, F., Luthra, R., Ye, Y., Wen, S., Berry, D.,
  and Kurzrock, R. (2012).
\newblock Personalized medicine in a phase {I} clinical trials program: the
  {MD} {A}nderson {C}ancer {C}enter initiative.
\newblock {\em Clin Cancer Res.}, 18(22):6373--83.

\bibitem[Tsimberidou, 2009]{impact}
Tsimberidou, A.~M. (2009).
\newblock Initiative for molecular profiling in advanced cancer therapy
  (impact) trial, an umbrella protocol.
\newblock \texttt{https://clinicaltrials.gov/}, study NCT00851032.
\newblock accessed, 02/09/2017.

\bibitem[Tsimberidou, 2014]{impact2}
Tsimberidou, A.~M. (2014).
\newblock {IMPACT} 2: {R}andomized study evaluating molecular profiling and
  targeted agents in metastatic cancer.
\newblock \texttt{https://clinicaltrials.gov/}, study NCT02152254.
\newblock accessed, 02/09/2017.

\bibitem[Tsimberidou et~al., 2014a]{Tismb&Schilsky:14}
Tsimberidou, A.~M., Eggermont, A.~M., and Schilsky, R.~L. (2014a).
\newblock {{P}recision cancer medicine: the future is now, only better}.
\newblock {\em Am Soc Clin Oncol Educ Book}, pages 61--69.

\bibitem[Tsimberidou et~al., 2014b]{TsimbAlBerry:14}
Tsimberidou, A.~M., Wen, S., Hong, D.~S., Wheler, J.~J., Falchook, G.~S., Fu,
  S., Piha-Paul, S., Naing, A., Janku, F., Aldape, K., Ye, Y., Kurzrock, R.,
  and Berry, D. (2014b).
\newblock {{P}ersonalized medicine for patients with advanced cancer in the
  phase {I} program at {M}{D} {A}nderson: validation and landmark analyses}.
\newblock {\em Clin. Cancer Res.}, 20(18):4827--4836.

\bibitem[Van~de Vijver et~al., 2002]{Vijver:2002}
Van~de Vijver, M., He, Y., van't Veer, L., Dai, H., Hart, A., Voskuil, D., and
  et~al. (2002).
\newblock {A gene-expression signature as a predictor of survival in breast
  cancer}.
\newblock {\em The New England Journal of Medicine}, 347:1999--2009.

\bibitem[Wathen and Thall, 2017]{WathenThall:17}
Wathen, J.~K. and Thall, P.~F. (2017).
\newblock A simulation study of outcome adaptive randomization in multi-arm
  clinical trials.
\newblock {\em Clinical Trials}, 14(5):432--440.

\bibitem[Xu et~al., 2013]{xu2013nonparametric}
Xu, Y., Lee, J., Yuan, Y., Mitra, R., Liang, S., M{\"u}ller, P., Ji, Y., et~al.
  (2013).
\newblock Nonparametric {B}ayesian bi-clustering for next generation sequencing
  count data.
\newblock {\em {B}ayesian Analysis}, 8(4):759--780.

\bibitem[Xu et~al., 2014]{xu2014subgroup}
Xu, Y., Trippa, L., M{\"u}ller, P., and Ji, Y. (2014).
\newblock Subgroup-based adaptive (suba) designs for multi-arm biomarker
  trials.
\newblock {\em Statistics in Biosciences}, pages 1--22.

\bibitem[Yang et~al., 2012]{yang2012antitumor}
Yang, H., Higgins, B., Kolinsky, K., Packman, K., Bradley, W.~D., Lee, R.~J.,
  Schostack, K., Simcox, M.~E., Kopetz, S., Heimbrook, D., et~al. (2012).
\newblock Antitumor activity of {BRAF} inhibitor vemurafenib in preclinical
  models of {BRAF}-mutant colorectal cancer.
\newblock {\em Cancer Research}, 72(3):779--789.

\bibitem[Zhou et~al., 2008]{battle:08}
Zhou, X., Liu, S., Kim, E.~S., Herbst, R.~S., and Lee, J.~J. (2008).
\newblock {{B}ayesian adaptive design for targeted therapy development in lung
  cancer--a step toward personalized medicine}.
\newblock {\em Clinical Trials}, 5(3):181--193.

\end{thebibliography}
\end{document}